\documentclass[a4paper]{jpconf}
\usepackage{amsmath,amssymb,bm}
\usepackage[dvips]{graphicx}
\usepackage{yfonts}
\usepackage{enumerate}
\newcommand{\beq}{\begin{eqnarray}}
\newcommand{\eeq}{\end{eqnarray}}

\newcommand{\SO}{\text{SO}}

\newcommand{\U}{\text{U}}

\begin{document}

\title{Some exact results on the QCD critical point}
\author{Yoshimasa Hidaka$^{1}$ and Naoki Yamamoto$^{2,3,4}$}
\address{
$^{1}$Theoretical Research Division, Nishina Center, RIKEN, Wako 351-0198, Japan\\
$^{2}$Yukawa Institute for Theoretical Physics, Kyoto University, Kyoto 606-8502, Japan\\
$^{3}$Institute for Nuclear Theory, University of Washington, 
Seattle, Washington 98195-1550, USA\\
$^{4}$Maryland Center for Fundamental Physics, Department of Physics, 
University of Maryland, College Park, MD 20742-4111, USA}
\ead{hidaka@riken.jp, nyama@umd.edu}

\begin{abstract}
We show, in a model-independent manner, that the QCD critical point can
appear only inside the pion condensation phase of the phase-quenched QCD 
as long as the contribution of flavor-disconnected diagrams is negligible.
The sign problem is known to be maximally severe in this region, implying
that the QCD critical point is reachable by the present lattice QCD techniques
only if there is an enhancement of the flavor-disconnected contribution at finite 
baryon chemical potential.
\end{abstract}

\section{Introduction and conclusion}
\label{sec:intro}
The phase structure of quantum chromodynamics (QCD) 
at finite temperature $T$ and finite baryon chemical potential $\mu_B$
has been a subject of considerable interest.
Among others, of particular interest is the possible QCD critical point,
which is the terminal point of the first-order 
chiral phase transition (for a review, see \cite{Stephanov:2004wx}).
Such critical points may appear not only in hot matter between the hadron phase
and quark-gluon plasma \cite{Asakawa:1989bq} but also in dense matter 
between the hadron phase and color superconductor
\cite{Kitazawa:2002bc, Hatsuda:2006ps}, and are potentially observable in
heavy-ion collision experiments at RHIC and FAIR.
Due to the sign problem, however, it has not yet been established in first principles 
lattice QCD simulations whether the QCD critical points really exist or not.

\begin{figure}[h]
\includegraphics[width=21pc]{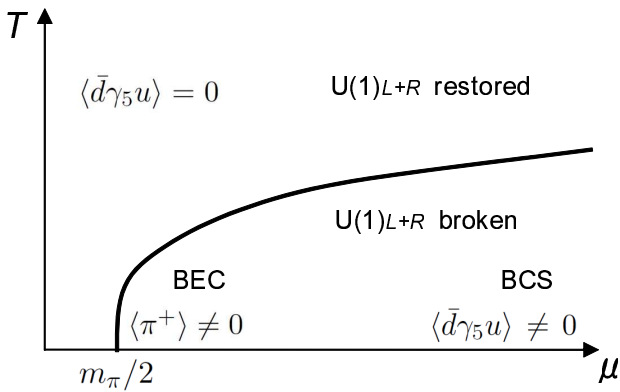}\hspace{2pc}%
\begin{minipage}[b]{14pc}\caption{\label{fig:muI}
Phase diagram of phase-quenched QCD with $\mu$ or 
QCD with isospin chemical potential $\mu_I=2\mu$ 
\cite{Son:2000xc} (figure taken from \cite{Hanada:2011ju}).
See the text in section~\ref{sec:muB} for further detail.}
\end{minipage}
\end{figure}

The first attempt to locate the QCD critical point was made by
Fodor and Katz based on the reweighting method \cite{Fodor:2001pe}. 
As pointed out by Splittorff \cite{Splittorff:2006vj}, however, their critical points 
happen to be located on the critical line of the phase-quenched QCD 
with the fermion determinant $\det {\cal D}$ replaced by $|\det {\cal D}|$
(see Figure~\ref{fig:muI} for the phase diagram),
where the reweighting method breaks down.
See also \cite{Ejiri:2005ts, Golterman:2006rw} for other but related problems.
Indeed in every model calculation (at least in the mean-field approximation), 
such as the random matrix model \cite{Han:2008xj}, 
Nambu-Jona-Lasinio (NJL) model \cite{Andersen:2009zm}, 
and Polyakov-loop extended NJL model \cite{Sakai:2010kx}, 
critical points are somehow observed only inside the critical line of the 
phase-quenched theory where the sign problem is maximally severe 
\cite{Splittorff:2006fu, Splittorff:2006vj}.
The question is whether this is the case in real QCD; 
if so, the QCD critical point is quite difficult to access by 
the present lattice QCD techniques.

In this article we show, in a model-independent manner, that the QCD critical point
cannot exist under the following conditions \cite{Hidaka:2011jj}:
\begin{enumerate}
\item{The contribution of flavor-disconnected diagrams is negligible
compared with that of connected diagrams.}
\item{The coordinate ($T,\mu$) in QCD at finite $\mu_B=3\mu$ 
is outside the critical line of the phase-quenched QCD.
(Note that it is \emph{not} a physical phase transition in QCD at finite $\mu_B$.)}
\end{enumerate}
The condition (i) is satisfied in the large-$N_c$ QCD and the models 
in the mean-field approximation, but it could be violated in real QCD. 
In other words, if the QCD critical point exists in a region accessible by the 
present lattice QCD techniques, the condition (i) must be broken;
it necessitates some kind of enhancement of the flavor-disconnected
contribution, which should also be observed in the lattice QCD simulations
(see section~\ref{sec:discussion} for more details).

\section{Theorem on the QCD critical point}
\subsection{Sketch of the proof}
\label{sec:sketch}
The main observation to arrive at the conclusion above is that pions, which are the 
(pseudo-)Nambu-Goldstone modes associated with chiral symmetry breaking, 
are the lightest among all the mesonic modes (even in the medium)
as long as the above conditions (i) and (ii) are met \cite{Hidaka:2011jj}.%
\footnote{It is interesting to note that the maximal severity of the sign problem 
inside the critical line of the phase-quenched QCD is also related to the lightness
of pions \cite{Grabowska:2012ik} (see also \cite{Splittorff:2006fu}).}
We empirically know that this is true in the QCD vacuum.
If this is also true in some region at finite $T$ and/or $\mu_B$, 
the flavor-singlet mesonic mode $\sigma$, which has the quantum number 
$\bar \psi \psi$ with the mass $m_{\sigma}$, satisfies the relation: 
\beq
\label{eq:bound}
m_{\sigma} \geq m_{\pi}>0,
\eeq
where $m_{\pi}$ is the pion mass and the latter inequality follows 
due to the explicit breaking of chiral symmetry in the presence of the quark mass.

On the other hand, the definition of the QCD critical point is that
the static correlation length $\xi$ defined by
\beq
\langle \bar \psi \psi ({\bm x}) \bar \psi \psi ({\bm y}) \rangle 
\sim e^{- |{\bm x}-{\bm y}|/\xi},
\eeq
is divergent, or $m_{\sigma}=1/\xi$ is vanishing at the point.
However, from the bound  (\ref{eq:bound}) for $m_{\sigma}$, 
this is apparently impossible, and thus, the QCD critical point cannot 
exist in this region.
Below we shall show, based on the rigorous QCD inequalities \cite{Weingarten:1983uj},
that pions are indeed the lightest mesonic modes under the conditions (i) and (ii) 
given in section~\ref{sec:intro}. We shall work in the Euclidean and two-flavor QCD
with the degenerate quark mass $m$. 
(For the generalization to QCD with any number of flavors $N_f$, see \cite{Hidaka:2011jj}.)

\subsection{Proof at $\mu_B=0$}
For clarity, we first illustrate the essence of the idea at $\mu_B=0$, 
and will generalize it to the case for $\mu_B \neq 0$.
Consider the Dirac operator ${\cal D} = D + m$, where 
the operator $D=\gamma_{\mu} (\partial_{\mu}+ig A_{\mu})$. 
satisfies the anti-Hermiticity and chiral symmetry, 
$D^{\dag}=-D$ and $\gamma_5 D \gamma_5 = -D$.
From these two properties, we can easily show
\beq
\label{eq:gamma5}
\gamma_5 {\cal D} \gamma_5 = {\cal D}^{\dag},
\eeq
and the positivity, $\det {\cal D} \geq 0$.

Let us now consider a correlation function for a generic flavor nonsinglet 
fermion bilinear $M_{\Gamma}=\bar \psi \Gamma \psi$, 
\beq
C_{\Gamma}(x, y) \equiv \langle M_{\Gamma}(x) M_{\Gamma}^{\dag}(y) \rangle_{\psi, A}
= - \langle \tr [S_A(x,y) \Gamma S_A(y,x) \bar \Gamma] \rangle_A.
\eeq
Here, $S_A(x,y) \equiv \langle x|{\cal D}^{-1} |y \rangle$ denotes 
a propagator from the point $y$ to $x$ in a background gauge field $A$,
the symbols $\langle \ \cdot \ \rangle_{\psi, A}$ and $\langle \ \cdot \ \rangle_{A}$
denote the average over $\psi, A$ and the average over $A$, respectively,
and $\bar \Gamma \equiv \gamma_0 \Gamma^{\dag} \gamma_0$.
From (\ref{eq:gamma5}) and the positivity, we have
\beq
\label{eq:inequality}
C_{\Gamma} = \langle \tr [S_A(x,y) \Gamma i\gamma_5 
S^{\dag}_A(x,y) i\gamma_5 \bar \Gamma] \rangle_A 
\leq \langle \tr [S_A(x,y) S^{\dag}_A(x,y)] \rangle_A.
\eeq
Here we used the Cauchy-Schwarz inequality, which is saturated when 
$\Gamma = i\gamma_5 \tau_A$ with $\tau_A$ being the traceless flavor generators.
The asymptotic behavior of $C_{\Gamma}$ at large distance $|{\bm x}-{\bm y}|$ 
in the static limit can be written as
\beq
C_{\Gamma} \sim e^{-m_{\Gamma}|{\bm x}-{\bm y}|},
\eeq
where $m_{\Gamma}$ is the mass of the lowest mesonic 
state in the channel $\Gamma$.
Then the inequality (\ref{eq:inequality}) leads to 
\beq
\label{eq:bound_general}
m_{\Gamma} \geq m_{\pi}.
\eeq

In this derivation of the QCD inequalities, we need to assume that 
$M_{\Gamma}$ is not flavor singlet, otherwise the contribution of 
flavor-disconnected diagrams
$\langle \tr[\Gamma S_A(x,x)]\tr[\bar \Gamma S_A(y,y)] \rangle_A$
must also be taken into account.
Phenomenologically disconnected diagrams may be suppressed compared 
with connected ones, as is known as the Okubo-Zweig-Iizuka (OZI) rule. 
Theoretically it can be justified in the 't Hooft large-$N_c$ limit 
\cite{'tHooft:1973jz}, since the flavor-disconnected diagrams are shown to be 
subleading in $1/N_c$ compared with the connected ones.
Here we simply assume that the contribution of the flavor-disconnected diagrams
is negligible [condition (i)], where (\ref{eq:bound_general}) is also applicable
to the flavor-singlet channel, $\Gamma=\sigma$.

This argument is valid for any temperature $T$.
Therefore, we conclude that the QCD critical point cannot exist 
on the $T$-axis (at $\mu_B=0$) under the condition (i).

\subsection{Proof at $\mu_B \neq 0$}
\label{sec:muB}
Unfortunately the above argument cannot be naively generalized to $\mu_B \neq 0$,
because the Dirac operator ${\cal D}(\mu) = D + \mu \gamma_0 + m$ does no
longer satisfy (\ref{eq:gamma5}) and the positivity of the measure is lost.
This is indeed the origin of the notorious sign problem in QCD at $\mu_B \neq 0$.

On the other hand, we can generalize it to the phase-quenched QCD, or equivalently,
QCD at finite isospin chemical potential $\mu_I = 2\mu$.
In this case, the Dirac operator ${\cal D}(\mu_I) = D + \mu_I \gamma_0 \tau_3/2 + m$
satisfies the relation $\tau_1 \gamma_5 {\cal D} \gamma_5 \tau_1 = {\cal D}^{\dag}$ 
and $\det{\cal D}(\mu_I) \geq 0$ \cite{Son:2000xc}; or complex Dirac eigenvalues 
for two flavors are conjugate with each other, and 
$\det{\cal D}(\mu_I) = |\det(\mu)| \geq 0$.
Then the following inequality follows at {\it any} $\mu_I$ \cite{Son:2000xc}:
\beq
C_{\Gamma} \leq \langle \tr [S_A(x,y) S^{\dag}_A(x,y)] \rangle_A,
\eeq
which is saturated when $\Gamma=i\gamma_5 \tau_{1,2}$.
This leads to the inequality, $m_{\Gamma} \geq m_{\pi_{\pm}}$. 
Under the condition (i) this is again applicable to $\Gamma = \sigma$, 
and repeating a similar argument to QCD at $\mu_B=0$, 
the critical point would be prohibited at $\mu_I \neq 0$.

However, one should note that this is not necessarily applicable to any coordinate
($T, \mu_I$). Indeed when the pion condensation phase appears (where $m_{\pm}=0$)
as we will see shortly, $m_{\sigma}=0$ is still possible and the QCD critical point 
is not ruled out.%
\footnote{The statement in \cite{Hidaka:2011jj} that the critical point cannot 
appear in QCD at any $\mu_I$ in the large-$N_c$ limit was not precise. However, 
the conclusion that the QCD critical point is ruled out at finite $\mu_B$ 
under the conditions (i) and (ii) remains unchanged.}

Let us look at the phase diagram of the phase-quenched QCD
(figure~\ref{fig:muI}), first discussed in \cite{Son:2000xc}.
As one increases $\mu_I$ at $T=0$, $\pi_+$ meson is first excited 
(instead of a baryon) so that it exhibits the Bose-Einstein condensation (BEC) 
for $\mu>m_{\pi}/2$. 
On the other hand, at large $\mu_I$ where fundamental degrees of freedom
are quarks and gluons, the attractive interaction between $u$ and $\bar d$ 
near the Fermi surface leads to the Bardeen-Cooper-Schrieffer (BCS) pairing 
$\langle \bar d \gamma_5 u \rangle$ \cite{Son:2000xc}. 
Because of the same symmetry breaking patterns $\U(1)_{L+R} \rightarrow {\mathbb Z}_2$, 
it is plausible that there is no phase transition between the two regimes. 
This is the so-called BEC-BCS crossover phenomena. 
Below we simply call it the pion condensation phase.
Combining the arguments above altogether, it follows that the critical point 
cannot appear in QCD at $\mu_I \neq 0$ outside the pion condensation phase 
under the condition (i).

Now we can generalize this result to QCD at $\mu_B \neq 0$,
taking into account the following property which holds 
under the conditions (i) and (ii):
\beq
\label{eq:equivalence}
\langle \bar \psi \psi \rangle_{\mu_B=3\mu}
= \langle \bar \psi \psi \rangle_{\mu_I=2\mu},
\eeq
where the same ($T,\mu$) is taken in both hand sides.
This property can be understood as follows \cite{Cohen:2004mw}:
the contribution of $u$ and $d$ to the chiral condensate are decoupled 
from each other under the condition (i), and there is a charge conjugation 
symmetry which flips the sign of the chemical potential for $u$ and $d$ independently.
As ($\mu_u, \mu_d$)=($\mu, \pm \mu$) correspond to the baryon and isospin 
chemical potentials respectively, the relation (\ref{eq:equivalence}) follows.
However, this is not true inside the pion condensation phase of the 
phase-quenched QCD where $u$ and $d$ are coupled in the ground state,
and hence, the condition (ii) is necessary \cite{Hanada:2011ju, Hanada:2012es}.
Therefore the locations and the orders of the chiral transition must be the same 
between QCD at $\mu_B \neq 0$ and phase-quenched QCD
outside the pion condensation phase of the latter.
The exactness of the phase quenching can be actually shown for \emph{any} observable 
to $O(N_f/N_c)$ \cite{Hanada:2011ju, Hanada:2012es} using the large-$N_c$ 
orbifold equivalence \cite{Bershadsky:1998cb} between QCD at $\mu_B$ 
and QCD at $\mu_I$ (see also \cite{Cherman:2010jj} for the orbifold 
equivalence between QCD and $\SO(2N_c)$ gauge theory at $\mu_B \neq 0$).

Since the QCD critical point is already shown to be excluded outside the pion 
condensation phase of the phase-quenched QCD, the same must be true for
QCD at $\mu_B \neq 0$. This is our main result \cite{Hidaka:2011jj}. 
This also provides the underlying reason why the critical points are always 
observed in this region with maximal severity of the sign problem 
in the model calculations \cite{Han:2008xj, Andersen:2009zm, Sakai:2010kx}.

\section{Discussion}
\label{sec:discussion}
So far we have assumed the condition (i), but how is this reasonable in real QCD? 
To understand it, first note that as far as we assume the condition (i), 
the chiral transition cannot be first order outside the pion condensation 
phase from our theorem. 
This is because, if it was first order, we could weaken the phase transition by 
gradually increase the quark mass $m$ and we would eventually arrive 
at the critical mass $m=m_c$ where the chiral transition becomes second order; 
this is prohibited from our theorem.
It means that the only thing that can make the chiral transition stronger to be 
first order is the effect of flavor-disconnected diagrams, 
including the effect of the axial anomaly \cite{Pisarski:1983ms}.
So whether the QCD critical point 
can appear outside the pion condensation phase is determined by the competition 
between the contribution of disconnected diagrams and the quark mass.
We know from the lattice QCD simulation \cite{Aoki:2006we} that the latter 
overwhelms the former at $\mu_B=0$ to make the chiral transition crossover
at finite $T$, and the condition (i) is satisfied in this sense.
Therefore, unless there is enhancement of the latter such that it overcomes the former 
at $\mu_B \neq 0$ for some reason, the critical point would not 
be accessible by the present lattice QCD techniques. 

In addition, nuclear matter is shown to lie inside this pion condensation phase 
independently of the quark mass, at least $T=0$ \cite{Cohen:2003ut}. 
Most likely these physics at finite baryon density appear due to the 
subtle cancellation of the complex phase.
It is certainly an important challenge for the theory to find out other ways to sample 
appropriate configurations in this region, which could enable us to 
probe the interesting physics in dense QCD including the color superconductivity.
(For such recent attempts, see, e.g., \cite{Aarts:2008wh, 
Cristoforetti:2012su, Grabowska:2012ik, Ohnishi:2012yb}.)

\ack
Y.H. was supported by JSPS KAKENHI Grant Numbers 23340067 and 24740184.
N.Y. is supported by JSPS Research Fellowships for Young Scientists.

\end{document}